\documentclass[aps, twocolumn,groupedaddress,nofootinbib,nobalancelastpage,nobibnotes]{revtex4}
\pdfoutput=1
\usepackage{amsmath,amsfonts,amssymb,mathrsfs,graphicx,color}
\usepackage[squaren]{SIunits}
\usepackage{verbatim}
\usepackage{enumerate}
\usepackage{graphicx}
\usepackage{ulem}
\usepackage[hyperfootnotes=false]{hyperref}
\usepackage{xcolor}
\usepackage{slashed}
\usepackage[utf8]{inputenc}
\usepackage[mathscr]{euscript}
\usepackage{hyperref}
\hypersetup{colorlinks=true, citecolor=red, linkcolor=blue, urlcolor=black}

\usepackage{comment,color}

\begin{document}

\title{Neutrinos and $\gamma$-rays  from the Galactic Center Region 
After H.E.S.S.\ Multi-TeV Measurements}

\author{F. Vissani}\email{francesco.vissani@lngs.infn.it}
\affiliation{INFN, Laboratori Nazionali del Gran Sasso, Assergi (AQ), Italy}
\affiliation{Gran Sasso Science Institute, L'Aquila (AQ), Italy}
\author{A. Palladino}\email{andrea.palladino@gssi.infn.it}
\affiliation{Gran Sasso Science Institute, L'Aquila (AQ), Italy}
\author{S. Celli}\email{silvia.celli@gssi.infn.it}
\affiliation{Gran Sasso Science Institute, L'Aquila (AQ), Italy}

\begin{abstract}
The hypothesis of a PeVatron in the Galactic Center, emerged with the recent $\gamma$-ray measurements of  
H.E.S.S.\ \cite{nature}, motivates the search for neutrinos  from this source. 
The effect of $\gamma$-ray absorption is studied:
at the energies currently probed, the 
known background radiation field{s} lead to small effects, whereas it is not possible to exclude large effects 
due to new IR radiation fields near the very Center. 
Precise upper limits on neutrino fluxes are derived and 
the underlying hypotheses are discussed. The expected number of events for ANTARES, IceCube and KM3NeT, {based on} the H.E.S.S.\ measurements, are calculated. It is shown that km$^3$-class telescopes in the Northern hemisphere  have the potential of  observing high-energy neutrinos from this 
important astronomical object and can check {the} existence of a hadronic PeV galactic accelerator.
\end{abstract}

\maketitle



\section{Introduction}
The supermassive black-hole in the center of the Milky Way, located in the radio source Sgr~A*, is one of the most interesting astronomical objects: see Ref.~\cite{smbh} for an extensive review. 
It is now in a state of relative inactivity~\cite{Ponti} but there is no good reason {for it to be} stationary. E.g., there are interesting hints for a much stronger emission few 100 years ago \cite{64g};   
on the time  scale of 40,000  years, major variability episodes are 
 expected~\cite{smbh2};
 Fermi bubbles \cite{fb} could be visible manifestations  
\cite{fba} of its intense activity.
 Therefore, it is reasonable to expect that a past emission from the 
  Galactic Center leads to 
observable effects.
Such scenario was recently considered in Ref.~\cite{fujita}.

The latest observations by the H.E.S.S.\ observatory \cite{nature},   that various regions
around Sgr~A* emit $\gamma$-rays 
till many tens of TeV, 
are offering us new occasions to investigate this object. These  $\gamma$-rays obey 
non-thermal distributions, which are moreover different in the closest vicinity of Sgr~A* and in its outskirts.
In the latter case, the $\gamma$-rays seem to extend till very high energies {($\sim 35$ TeV)} without a perceivable cut-off.
  
The $\gamma$-rays seen by H.E.S.S.\ can be attributed to cosmic ray collisions \cite{nature}. This is a likely hypothesis, but the proof of its correctness requires neutrino telescopes. In this connection, it is essential to derive reliable predictions for the search of a neutrino signal from Sgr~A* and its surroundings, and H.E.S.S.\ observations are very valuable in this respect.  Remarkably, the possibility that the Galactic Centre is a significant neutrino source is discussed since the first works~\cite{zhe} and it is largely within expectations: indeed Sgr~A* is one of the main point source targets, already for the IceCube observatory~\cite{icnew}.

In this work, we discuss the implications of  the findings of H.E.S.S.,  briefly reviewed in Sect.~\ref{sec:gammaRay}, 
{where we also explain our assumptions on the 
$\gamma$-ray spectra at the source.}
The effect of $\gamma$-ray absorption (due to the 
known radiation fields or to new ones, close to the Galactic Center) 
is examined in details in
Sect.~\ref{sec:abs}. 
The expected signal in neutrino telescopes,   
evaluated at the best of the present knowledge, {is shown in Sect.~\ref{sec:ripu} and it is quantified in 
Sect.~\ref{sec:rates}}, while Sect.~\ref{sec:ccc} is devoted for the conclusion.
We argue that the PeVatron hypothesis makes the case for a cubic kilometer class neutrino telescope, located in the Northern hemisphere, more compelling than ever. 

\section{The  $\gamma$-ray spectra from the Galactic Center Region\label{sec:gammaRay}}

The excess of VHE $\gamma$-rays reported by the H.E.S.S.\ collaboration \cite{nature} comes from two regions around the Galactic Center: a Point Source (HESS J1745-290), identified by a circular region centered on the {radio source} Sgr~A* with a radius of 0.1$^{\circ}$,  and a Diffuse emission, coming from an annulus with inner and outer radii of 0.15$^{\circ}$ and 0.45$^{\circ}$ respectively. The observed spectrum from the Point Source is described by a cut-off power law
distribution, as
\begin{equation}
\phi_\gamma(E)=\phi_0 \left( \frac{E}{1 \, \textrm{TeV}} \right)^{-\Gamma} \exp\!\left( {-\frac{E}{E^{\gamma}_{\textrm{\tiny cut}}}} \right)
\label{eq:plcut}
\end{equation} 
while in the case of Diffuse emission an unbroken power law is preferred; in the last case, however, also cut-off power law fits are  presented, as expected from standard mechanisms of particle acceleration into the Galaxy.

 {The H.E.S.S.\ collaboration has summarised {its} observations  by means of the following parameter sets:}
\begin{itemize}
\item Best fit of the Point Source (PS) region: \\ $\Gamma=2.14 \pm 0.10$,\\ $\phi_0=(2.55 \pm 0.37) \times 10^{-12}$ TeV$^{-1}$ cm$^{-2}$ s$^{-1}$,\\ $E^{\gamma}_{\textrm{\tiny cut}}=10.7 \pm 2.9$ TeV;
\item Best fit of the Diffuse (D) region: \\ $\Gamma=2.32 \pm 0.12$,\\ $\phi_0=(1.92 \pm 0.29) \times 10^{-12}$ TeV$^{-1}$ cm$^{-2}$ s$^{-1}$;
\end{itemize}
The best-fits of both the Diffuse and the Point Source emission are shown in Fig.~\ref{figabs}, right panel. 

However, in order to predict the  neutrino spectrum, the $\gamma$-ray spectrum at the source--i.e. the emission spectrum--is needed.
{We will discuss the implication of the assumption that the  emitted spectra coincide with the observed spectra as described by the previous functional forms and furthermore 
we will discuss the assumption that the $\gamma$-ray emission at the source is 
described by different model parameters,  namely:}
\begin{itemize}
\item Point Source emission with an increased value of the cut-off (PS*): \\ $\Gamma=2.14$,\\ $\phi_0=2.55 \times 10^{-12}$ TeV$^{-1}$ cm$^{-2}$ s$^{-1}$,\\ $E^{\gamma}_{\textrm{\tiny cut}}=100$ TeV;
\item Diffuse emission as a cut-off (DC) power law with: \\ $\Gamma=2.32$,\\ $\phi_0=1.92 \times 10^{-12}$ TeV$^{-1}$ cm$^{-2}$ s$^{-1}$, \\ $E^{\gamma}_{\textrm{\tiny cut}}=0.4 \,  \textrm{PeV}, 0.6  \,  \textrm{PeV}\,  \textrm{or} \,2.9$ PeV.
\end{itemize}
{The interest in considering an {\em increased} value of the cut-off (the case PS*), 
that is the only case that differs significantly from the spectra observed by H.E.S.S., is 
motivated in the next section.
Instead, 
the inclusion of a cut-off for the emission from the Diffuse region
agrees with the observations of H.E.S.S.\ and 
is motivated simply by the expectation 
of a maximum energy available for particle acceleration.}


Note that the $\gamma$-ray observations  extend till 20-40 TeV. This is an important region of energy but it covers only the lower region that is relevant for neutrinos: the latter one extends till 100 TeV, as clear e.g., from Fig.2  and 3 of~\cite{exa} and 
Fig.1 of \cite{vissanga}.
In other words, it should be kept in mind that until $\gamma$-ray observations till few 100 TeV will become available--thanks to future measurements by HAWC \cite{hawc} and CTA \cite{cta}--the expectations for neutrinos will rely in part on  extrapolation and/or on theoretical modeling. In this work, unless stated otherwise, 
we rely on a `minimal extrapolation', {assuming that the above functional descriptions of the $\gamma$-ray spectrum are valid descriptions of the emission spectrum}.

A precise upper limit on the expected neutrino flux can be determined
from the H.E.S.S.\ measurement, assuming 
a  hadronic origin of the observed $\gamma$-rays.
The presence of a significant leptonic component of the $\gamma$-rays 
would imply a smaller neutrino flux.
In principle, however, also other regions close the the Galactic Center, but not probed by H.E.S.S.,
could emit high-energy $\gamma$-rays and neutrino radiation, leading to an interesting signal.
{One reason is that the annulus, chosen by H.E.S.S.\ for the analysis, 
resembles more a region selected for observational purposes rather than an object with 
an evident physical meaning;\footnote{Again because of this consideration, and also in view of the fact that the angular resolution of the neutrino telescopes operated in water matches the physical size of the two regions, we will present  
the predictions for the Point Source and the Diffuse region separately.}
another reason is that 
the ice-based neutrino telescope  IceCube   integrates on an
angular extension of about $1^\circ$, which is 5 times larger than the 
angular region covered in \cite{nature}.}
In view of these reasons, the theoretical 
upper limit on the neutrino flux that we will derive is the {\em minimum} that is justified by the current $\gamma$-ray data.
Moreover, there is also a specific phenomenon that 
increases the expected neutrino flux that can be derived from the $\gamma$-ray flux currently measured by H.E.S.S.:
{this is the absorption of $\gamma$-rays from non standard radiation fields,} as discussed in the next section.


\section{Absorption of $\gamma$-rays}
\label{sec:abs}

During their  propagation in the  background radiation fields of the Milky Way, high-energy  photons are subject to absorption.
   Consider the {\em observed} $\gamma$-ray spectrum, as summarized by means of a certain functional form.
   The corresponding {\em emission} spectrum is larger: this is obtained modeling and then removing the effect of absorption (de-absorption).  
  The  neutrino spectrum corresponds to the emission spectrum, and thus it is larger than the one obtained by converting 
  the observed {$\gamma$-ray} spectrum instead. {Note that the idea that the $\gamma$-rays could suffer significant absorption already at H.E.S.S.\ energies  
was put forward in Ref.~\cite{nature}; here, we examine it in details.}

\paragraph{Description of the procedure}

The existence of a  Cosmic Microwave Background (CMB), 
that pervades the whole space and it 
is uniformly distributed, is universally known; this leads to absorption of
$\gamma$-rays of very high energies, around PeV. 
For what concerns the interstellar radiation background,  
the model   by Porter  et al.~\cite{Porter:2008ve},
  adopted e.g., in the   GALPROP simulation program \cite{galprop}, 
  can be conveniently used to  describe $\gamma$-ray absorption due to    
   {the InfraRed (IR) and StarLight (SL) backgrounds}
  (see e.g., \cite{moska,Lee:1996fp,cirelli,arman}), that occurs at lower energies.
  
  {It is convenient to group these three radiation fields (for instance CMB, IR and SL) as `known' 
  radiation fields, since it is not possible to exclude that in the vicinity of the Galactic Center 
  new intense IR fields exist, and thus we should be ready to consider also hypothetical or  `unknown' radiation fields.}
  The formal description of their absorption effects 
  can be simplified without significant loss of accuracy  if 
  the radiation  background field is effectively parameterized 
  in terms of a sum of thermal  and quasi-thermal distributions, 
  where the latter ones are just proportional to thermal distributions.  
  
For the $i$-th component of the radiation background, 
  two parameters are introduced: the temperature $T_i$ and the 
  coefficient of proportionality to the thermal distribution, that we call
  the `non-thermic  parameter' and  that we denote  by $\xi_i$.
  The 
 reliability of this procedure for the description of the Galactic absorption was already tested in \cite{Lee:1996fp,cirelli}.
  We found that the formalism
   can be simplified even further 
  without significant loss of accuracy thanks to a fully analytical (albeit approximate) formula,  derived and discussed in details in the appendix.  We have checked the excellent consistency with the other  approach--based on~\cite{Porter:2008ve}--by comparing our results with Fig.3    of~\cite{arman}.

   We emphasize a few advantages of this procedure, \\
 1)~the results are exact in the case of the CMB
 distribution, that is a thermal distribution;\\
 2)~such a procedure allows one   
  to vary the parameters of the radiation field very easily,
   discussing the effect of errors and uncertainties;\\
   3) the very same formalism allow us to model the effect of new hypothetical radiation background.

\begin{figure*}[t]
\centerline{\includegraphics[width=0.42\textwidth,angle=0]{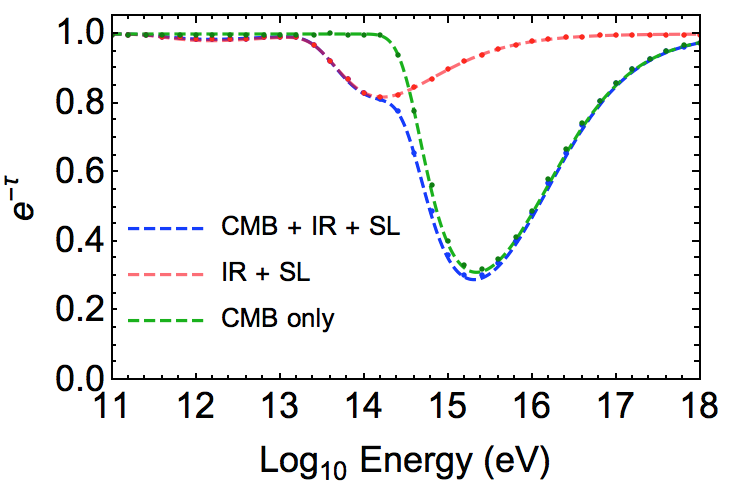}\hfill
\includegraphics[width=0.47\textwidth,angle=0]{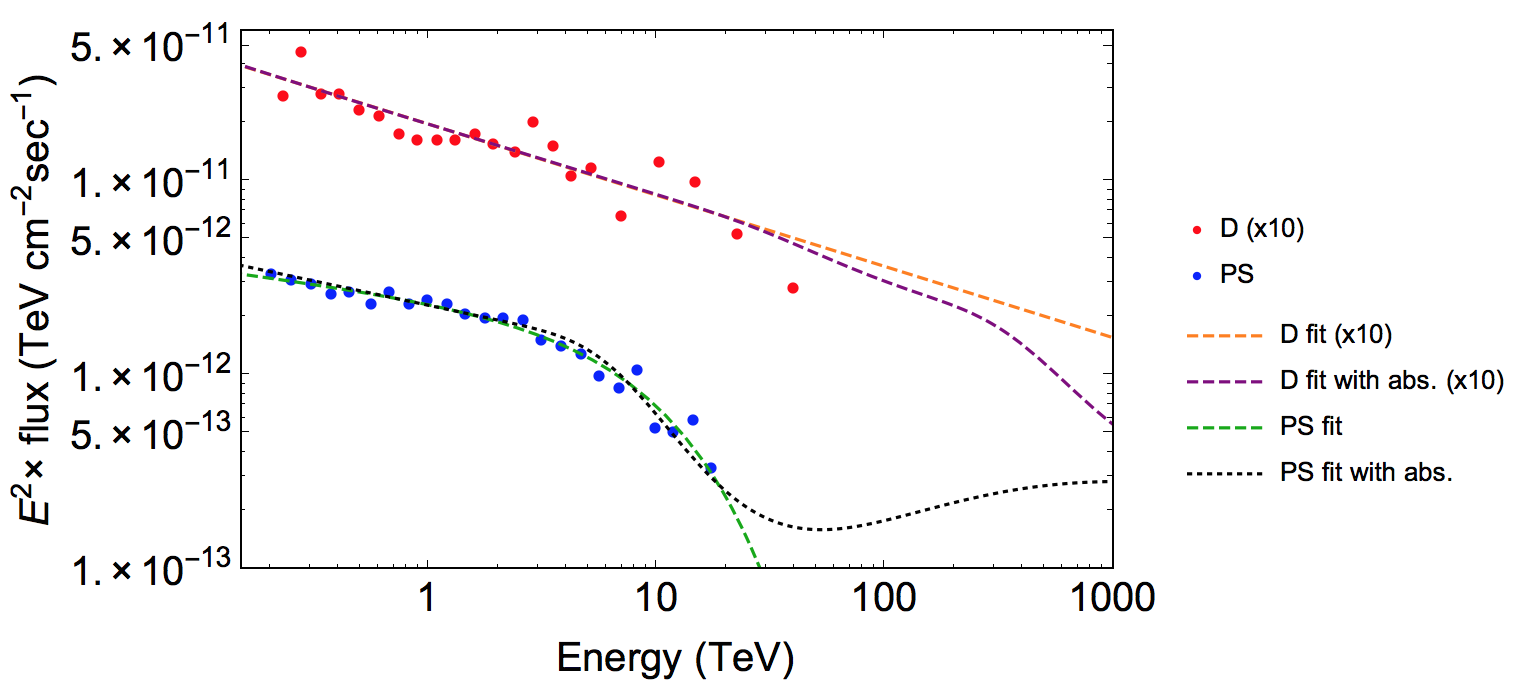}}
\caption{{\em Left panel:} Absorption of $\gamma$-rays from the Galactic Center at different energies, due to the interaction with CMB, IR and SL. {\em Right panel:} fits of the H.E.S.S.\ data.
D is the Diffuse flux, PS is the Point Source flux. For D, absorption due to CMB, IR and SL is considered. For PS, an increased  absorption, due to non standard radiation field, is considered. 
{As visual aid, in the right panel  we indicate  
the central values of the measurements for the Diffuse and the Point 
Source by red and blue dots, respectively, as obtained in ref.~\cite{nature}, where uncertainties can also be found.}}
\label{figabs}
\end{figure*}

\paragraph{Formalism}
A photon with energy $E_\gamma$ emitted from an astrophysical source can interact during its travel to  the Earth with ambient photons, producing electron-positron pairs. 
The probability that it will reach the Earth is 
\begin{equation}
P(E_\gamma)=\exp\left[  -\tau(E_\gamma) \right]
\end{equation}
where $\tau$ is the opacity.
In the interstellar medium, different radiation fields can offer a target to astrophysical photons and determine their absorption: the total opacity is therefore
the sum of various contributions,  
\begin{equation}
\tau=\sum_i\tau_i
\end{equation}
where the index $i$ indicates the components of the background radiation field that causes absorption. These include the CMB, as well as the IR and SL backgrounds, and possibly new ones, present near the region of the 
Galactic Center.

\begin{table}[t]
\caption{Values of the parameters of the background radiation fields used for the computation of the absorption factor of the $\gamma$-rays from the Galactic Center: the black body temperature $T_i$, the non-thermic parameter $\xi_i$,  the typical length $L_i$
{(namely the distance of the Galactic Center for CMB and 
the exponential scales for the IR and SL radiation fields)} 
the total density of photons $n_{\gamma,i}$ 
(obtained from Eq.~\ref{nonno}), 
the typical energy $E_i$ 
(obtained from Eq.~\ref{nonna}).}
\begin{center}
\begin{tabular}{lccccc}
\hline
Rad.\ field & $T_i$ & $\xi_i$ & $L_i$& $n_{\gamma,i}$ & $E_i$ \\ 
& (eV) &  & (kpc) & (cm$^{-3}$) & (TeV) \\ 
\hline 
CMB &$2.35  \cdot 10^{-4}$  & 1 & 8.3 & 410.7   & 1111 \\ 
IR &$3.10 \cdot 10^{-3}$ & $ 1.55 \cdot 10^{-4}$ &4.1 &146.0  & 84.23  	\\
SL  &$3.44 \cdot 10^{-1}$  & $1.47  \cdot 10^{-11}$ &2.4 & 19.0  & 0.26  \\
\hline
\end{tabular}
\end{center}
\label{alt}
\end{table}

For a thermal distribution (or for a distribution proportional to a thermal distribution)
the opacity from the $i$-th component is given simply by 
\begin{equation}
 \tau_i(E_\gamma)=1.315 \times  \frac{L_i}{L_0} \times \frac{n_{\gamma,i}}{n_
 {\gamma,\mbox{\tiny CMB}}} \times f\!\left( \frac{E_i}{E_\gamma} \right) 
 \label{tau}
\end{equation}
\normalsize
where the quantities chosen for the 
normalization are 
$L_0$=10 kpc (a typical galactic distance) and  $n_
 {\gamma,\mbox{\tiny CMB}}=\rm 410.7 \ cm^{-3}$.
 The various quantities in this formula are discussed in details below,
the numerical values (adopted in the calculation) are given in Tab.~\ref{alt}.
  The  overall figure gathers few constants from the 
 thermal CMB distribution, from the interaction cross section, and  the distance $L_0$. Its expression is 
 \begin{equation}
 \frac{\pi }{2\ \zeta(3)} \times \displaystyle L_0\: r_e^2 \times n_
 {\gamma,\mbox{\tiny CMB}}=1.315
 \end{equation}
 where $r_e=e^2/(m_e c^2)\approx 2.818\times 10^{-13}$ cm is the classical electron radius and $\zeta(3)\approx 1.20206$ the 
  Riemann's zeta function.  
  {Here} we examine and discuss the various quantities appearing 
in Eq.~\ref{tau}.\\
\textit{i)} The parameter $L_i$ is the size of the background radiation field.
 In the case of the CMB, this is just the 
 {distance between the 
 Galactic Center and the detector (i.e., 8.3 kpc)   
 because the CMB is uniformly distributed throughout the  interstellar medium.}
 The  IR and SL radiation fields, instead,   
 obey an approximate exponential distribution from the Galactic Center. 
 {The density of photon is,
${n_{\gamma,i}}(L)={n_{\gamma,i}} \, e^{-L/L_i}$. 
Thus,
the product ${n_{\gamma,i}}\times L_i$ is the column density of photons, and also for the IR and SL fields (as for the CMB)  $L_i$ 
represents the effective size of the region where the 
radiation is present.}
The values of the scales $L_i$ are given in Tab.~\ref{alt}. \\
\textit{ii)} 
$n_{\gamma,i}$ is the total number of photons of the considered background radiation field. Comparing IR and SL with the CMB we found that the total number of photons, in the first two cases, are given by the following expression: 
\begin{equation}
n_{\gamma,i}(\xi_i, T_i)=\xi_i \times n_{\gamma,\mbox{\tiny CMB}} \times \left(\frac{T_i}{T_{\mbox{\tiny CMB}}}\right)^3
\label{nonno}
\end{equation}
where $\xi_i \le 1$ is a numerical factor, the
non-thermic parameter,
chosen to reproduce the observed
 distribution of ambient photons (the case $\xi_i=1$ is the thermal one). \\ 
\textit{iii)} The energy $E_i$ is linked to the black body (or quasi black body for IR and SL) temperature $T_i$, through the relation 
\begin{equation}
E_i(T_i)=\frac{m_e^2}{T_i}
\label{nonna}
\end{equation}
where $m_e$ is the electron mass and the assumed 
distribution is proportional to a thermal distribution with
temperature $T_i$. \\
\textit{iv)} 
The adimensional function $f(x)$, 
where $x=\frac{E_i}{E_\gamma}$, describes  how the absorption varies with the energy of the $\gamma$-ray.
It was first obtained in Ref.~\cite{moska} using 
the results of \cite{breit}. This is discussed in the appendix, 
where precise values are  
obtained by means of numerical calculation. 
We derived a simple approximated expression for such a function, 
\begin{equation}
f(x) \simeq -a \cdot x\cdot \log[1-\exp(-b\ x^c) ], \ \ \ 
\left\{
\begin{array}{l}
a=3.68\\
b=1.52\\
c=0.89
\end{array}
\right.
\label{eq:analyticApp}
\end{equation}
which is very easy to use and accurate to  within 3\%.

The values of $T_i$, $\xi_i$ and $L_i$ for the known components 
were found fitting the energy spectra of radiation reported in GALPROP (Ref.~\cite{galprop}).
They are summarized in Tab.~\ref{alt} for the CMB, IR and SL radiation fields from the Galactic Center to the Earth. The latter two contributions affect the survival probability at energies smaller than those due to the CMB photons.  {The parameters $n_{\gamma,i}$ and $E_i$ are also given in table Tab.~\ref{alt}, to allow one to use directly of Eq.~(\ref{tau}) just replacing the appropriate numerical values.}
In this formalism, a population of  quasi-thermal 
background photons  is characterized by the
parameters $T_i,\xi_i,L_i$, and it will yield the opacity factor  $\tau_i=\tau(E_\gamma;T_i, \xi_i \times L_i)$. Note that $\xi_i$ and $L_i$  
appear in Eq.~\ref{tau} only through their product, so for each component of the background radiation field (known or hypothetical)
we have a  {\em two parameter} description of photon absorption.

\paragraph{Results}
The {effects of absorption due to the known radiation fields of Tab.~\ref{alt}, that concern} the $\gamma$-rays propagating from the Galactic Center to the Earth, are illustrated in the left panel of Fig.~\ref{figabs}. They become relevant at some hundreds of TeV,  so  the $\gamma$-rays presently observed by H.E.S.S.\ from the Diffuse region  (and the models D and DC introduced above) are not significantly influenced by this phenomenon, 
assuming only the radiation fields~\cite{Porter:2008ve} used in GALPROP.
Therefore it is possible to use directly the observed diffuse $\gamma$-ray flux in order to obtain the $\gamma$-ray flux at the source,
{modulo the caveats concerning the extrapolation at high energy, see Sect.~\ref{sec:gammaRay}.}

On the other side, the flux from the Point Source  could be affected by the absorption due to a new,  non standard and intense radiation field close to the central black-hole.
{In fact, the physical conditions in the close vicinity of the supermassive black-hole are not perfectly known and in particular the local radiation field is not probed directly.}
Recall that the $\gamma$-ray spectrum from the Point Source {\em observed} by H.E.S.S.\ deviates from a power law distribution at the higher energies {currently probed: it would naturally differ from the} {\em emitted} $\gamma$-ray spectrum
in presence of a new radiation field.
{In  particular, 
the cut-off of the emitted spectrum would move towards higher energies.\footnote{{ 
Note incidentally that the position of the cut-off of the cosmic rays, that we suppose to be accelerated near the supermassive black-hole, is to date unknown and has to be deduced from the observations.}}}

We show below that,  hypothesizing a significant absorption of $\gamma$-rays close to the black-hole,  
the flux emitted from the Point Source is compatible with a power law distribution at the energies observed by H.E.S.S.\ and above,
{just as the one due to the Diffuse $\gamma$-ray component. This  speculative scenario allows us to estimate in a reasonable way the maximum effect of absorption due to yet unknown radiation fields.
Note that this has a direct implication on the neutrino signal, that we quantify later on.

In this scenario,} the background radiation field that causes absorption 
is characterized by a temperature $T=1.3 \times 10^{-2}$ eV, namely, should be in the far infrared spectral band. 
Hypothesizing a black body field, the corresponding density of photons is equal to $n^{\mbox{\tiny max}}_\gamma=7.04 \times 10^7 \ \rm cm^{-3}$, with a typical scale of the radiation field of $L_{\mbox{\tiny min}}=0.07 \rm \ pc$, namely with a column density corresponding to $L \times n_\gamma \simeq 5000 \ \rm \textrm{kpc cm}^{-3}$. If the radiation field is not exactly thermal, the number {density} of photons decreases whereas the typical length increases: for example, with a non-thermic parameter $\xi=0.02$, we obtain $n_\gamma=1.4 \times 10^6 \ \rm \textrm{cm}^{-3}$ and a typical length $L =3.5 \ \rm pc$. 

   This is illustrated in the right panel of Fig.~\ref{figabs}. The curve called ``PS fit with non standard abs.'' {is a power law spectrum with the same spectral index and normalization of the PS model of Sect.~\ref{sec:gammaRay}, that is modified taking into account the above scenario for $\gamma$-ray absorption}.

Some remarks on this scenario are in order:\\ 
{1)~Such non standard IR radiation field could be produced in the reprocessing of the radiation from the central source, due to collision with CircumNuclear Disk clumps, as reported in Ref.~\cite{smbh}.}\\
2)~Evidently, the Diffuse component would be not affected by this new IR radiation, because it is far enough from the Galactic Center.\\
3)~As one understands from Fig.~\ref{figabs}, if one wants to determine observationally whether 
the cut-off is intrinsic to the source or it is an absorption feature, measurements of $\gamma$-rays at energies above tens of TeV are required: indeed, the absorption results in a peculiar distortion of the power-law spectrum, that is expected to be different from the effect of an exponential cut-off above $\sim$50 TeV.  This discrimination should be possible with  CTA~\cite{cta} or with other future instruments.



\section{High-energy neutrinos from the Galactic Center Region\label{sec:ripu}}


Neutrinos could be produced in hadronic interactions of PeV protons with the ambient gas: since each neutrino carries about 5\% of the energy of the parent proton, we expect to see neutrinos in the multi-TeV range, in angular correlation with the high-energy $\gamma$-rays emitted from the Galactic Center Region. This scenario is supported by the observed correlation between the $\gamma$-ray emission and molecular clouds reported in \cite{nature}.

\paragraph{{Present upper limit on  neutrinos from Sgr~A*}}
To date, IceCube has set the best 90\% C.L. upper limit on the $\nu_\mu+\bar{\nu}_\mu$ flux assuming an unbroken E$^{-2}$ spectrum from Sgr~A* at \cite{icnew},
\begin{equation}
\phi_{\nu_\mu}+
\phi_{\bar{\nu}_\mu}=7.6 \times 10^{-12} \left( \frac{E}{\mbox{TeV}}\right)^{-2} \mbox{ TeV$^{-1}$ cm$^{-2}$ s$^{-1}$}  
\label{ict}
\end{equation}
Such a limit corresponds to the absence of a significant {event excess} over 
the known background, that in the analysis of IceCube  \cite{baka} amounts to 25.2 background events in a circle of $1^\circ$. {This limit has been obtained by means of downward-going track-type events, as discussed in the next section.} 

Presumably, this is the safest information we have on the neutrino emission from Sgr~A*, {even if it does not correspond to a  realistic assumption on the emitted neutrino spectrum. In principle, the assumption of a differential neutrino spectrum in the form of an E$^{-2}$ dependence would be a consequence of the first order Fermi acceleration mechanism, but there is no observational evidence that this is a reliable assumption, and moreover, this is not 
supported by the observed $\gamma$-ray spectrum}.

\begin{figure}[t]
\includegraphics[scale=0.49]{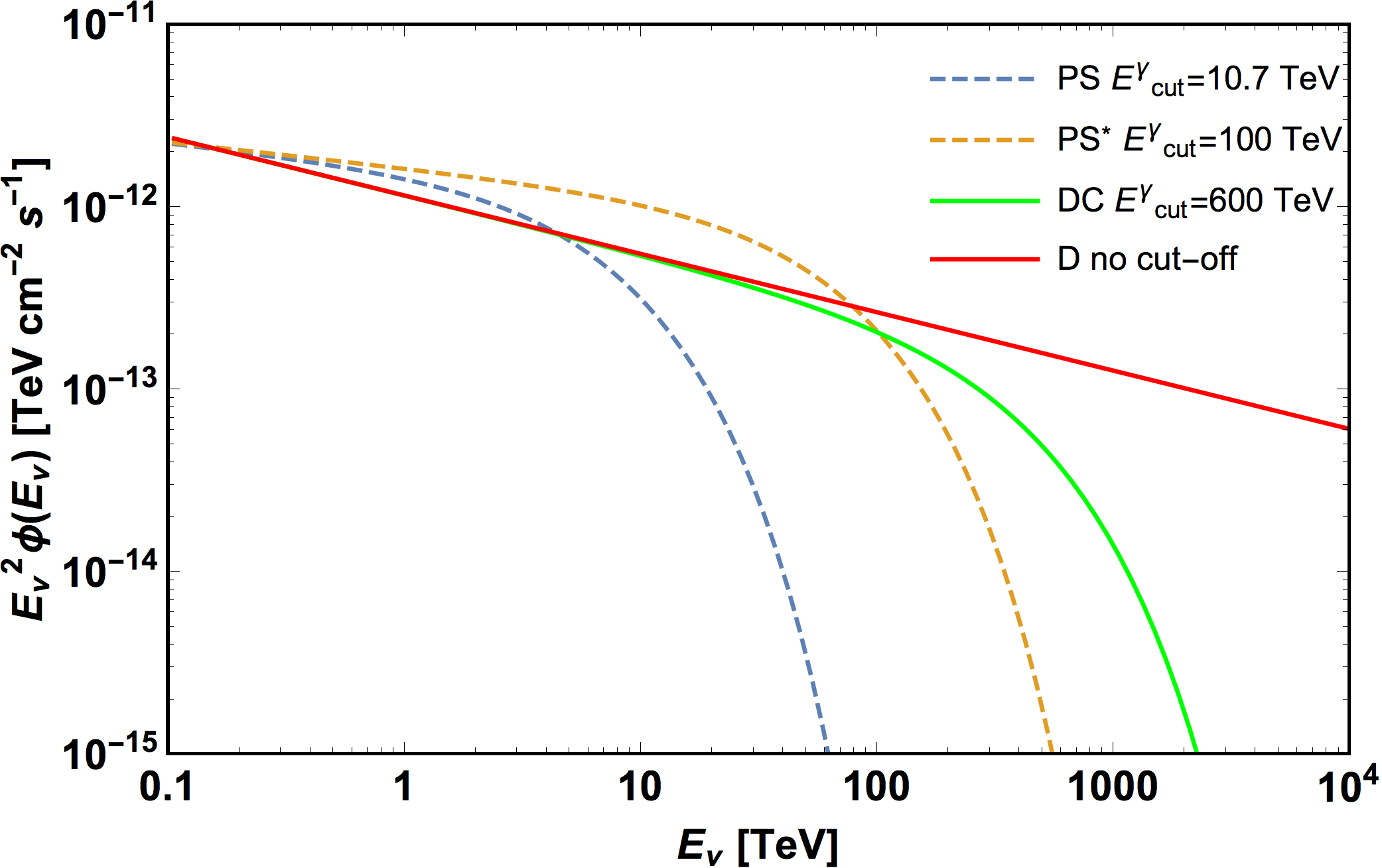}
\caption{{Predicted muon neutrino and antineutrino fluxes (summed) for the Point Source best fit with {a cut-off at E$^{\gamma}_{\textrm{\tiny cut}}=10.7$ TeV} and for an arbitrary cut-off at  {E$^{\gamma}_{\textrm{\tiny cut}}=100$} TeV. Also shown the fluxes for the Diffuse best fit without cut-off and with a cut-off at E$^{\gamma}_{\textrm{\tiny cut}}=600$~TeV. } }
\label{fig:fluxes}
\end{figure}

{
\paragraph{{Model prediction  / theoretical upper bound}}
The model prediction 
that we are going to derive is based instead on the current $\gamma$-ray observations and on the assumption that such emission is fully hadronic. The expectation 
that we obtain is well compatible with the IceCube non-detection. Indeed, the flux of  Eq.~\ref{ict} 
is much larger and has a distribution harder than 
the upper limit on the neutrino flux derived from the $\gamma$-ray observations.  This is evident from Fig.~\ref{fig:fluxes}, discussed in details just below.}

{Keeping in mind the crucial hypothesis, that the $\gamma$-rays observed by H.E.S.S.\ are {fully} hadronic, our model prediction can be regarded also as a theoretical upper bound.}
It is very important however to distinguish clearly the experimental upper limit of Eq.~\ref{ict} 
from this theoretical upper limit
on the expected neutrino signal,  
derived through $\gamma$-rays. The latter is more realistic and also 
much more stringent,  but, just as the former, it depends upon  
various theoretical assumptions.

In order to illustrate this point, we remark that the  $\gamma$-ray data collected by H.E.S.S.\ cannot exclude that  the $\nu_\mu+\bar\nu_\mu$ spectrum hardens to E$^{-2}$ above 20-40 TeV; however, the  normalization of this component has to be some 5 times smaller than Eq.~\ref{ict}, if the spectrum is a smooth distribution (a continuous function) linked to the observed $\gamma$-rays spectrum.
This kind of (very speculative) scenario, along with other scenarios mentioned in  
Sect.~\ref{sec:gammaRay}, might increase the expected neutrino signal. 

However, below in this work, we prefer   to 
focus conservatively 
on the minimal scenario that was defined in Sect.~\ref{sec:gammaRay} {as} it is motivated by H.E.S.S.\ 
measurements. 
{We will show that the theoretical limit on the neutrino flux,  corresponding to the $\gamma$-ray flux observed by H.E.S.S., is below the capabilities  of the detectors currently in data-taking, whereas it could be within the reach of the future  detectors.}

\paragraph{Method to calculate the muon neutrino flux}
Assuming a pu\-re\-ly hadronic origin of the emission $\gamma$-ray spectrum 
$\phi_{\gamma} (E)$, 
we can calculate the muon neutrino and antineutrino spectrum through the precise relations based on the assumption of 
cosmic ray-gas collisions  \cite{villantevissani},
\begin{equation}
\label{eq:Inu}
\begin{array}{rl}
\phi_{\stackrel{(-)}{\nu}_{\!\!\mu}}(E) & \displaystyle
=\alpha_\pi\ \phi_{\gamma} \left( \displaystyle  \frac{E}{1-r_\pi} \right)+\alpha_K\ \phi_{\gamma} \left( \displaystyle  \frac{E}{1-r_K} \right)+\\
& \displaystyle +\int_0^1{\frac{dx}{x}\ K_{\stackrel{(-)}{\nu}_{\!\!\mu}} (x)\  \phi_{\gamma}\! \left(\frac{E}{x} \right)}
\end{array}
\end{equation}
where $\alpha_\pi=0.380\ (0.278)$ and 
$\alpha_K=0.013\ (0.009)$ for $\nu_\mu$ and $\bar\nu_\mu$ respectively and 
%
where $r_x=(m_\mu/m_x)^2$ with $x=\pi,K$. In each expression, the first two contributions describe neutrinos from the two-body decay by pions and kaons, while the third term accounts for neutrinos from muon decay. The kernels for muon neutrinos $K_{\nu_\mu}(x)$ and for muon antineutrinos $K_{\bar{\nu}_\mu}(x)$, which also account  for oscillations from the source to the Earth, are
$$
\footnotesize{
K_{\nu_\mu}(x)\!=\! \left\{ 
\begin{array}{ll}
x^2(15.34-28.93x) & 0<x \leq r_K \\
0.0165+0.1193x+3.747x^2-3.981x^3 & r_K<x < r_\pi \\
(1-x)^2(-0.6698+6.588x) & r_\pi \leq x<1 \\
\end{array} 
\right.
}
$$
$$
\footnotesize{
K_{\bar{\nu}_\mu}(x)\!=\!  \left\{ 
\begin{array}{ll}
x^2(18.48-25.33x) & 0<x \leq r_K \\
0.0251+0.0826x+3.697x^2-3.548x^3 & r_K<x < r_\pi \\
(1-x)^2(0.0351+5.864x) & r_\pi \leq x<1 \\
\end{array} 
\right.
}
$$
Applying such procedure, the expected (upper limits on the)
neutrino spectra are obtained from the $\gamma$-ray spectrum. This is the closest we can go to a model-independent approach.

\paragraph{Results for the $\nu_\mu+\bar\nu_\mu$ fluxes}
We 
show in Fig.~\ref{fig:fluxes} the sum of the muon neutrino and antineutrino fluxes, derived using {for $\phi_{\gamma} (E)$} the 
four models introduced in Sect.~\ref{sec:gammaRay}, namely: 
1)~the best fit flux of the Point Source region (with 10.7~TeV cut-off), 
2)~the same one assuming that  the cut-off is at 100 TeV, 
3)~the 
best fit flux of the Diffuse region (without cut-off), 
4)~the same one including a cut-off at 600~TeV.
\newline
\indent
Our results compare  reasonably well with the fluxes gi\-ven  
in the Extended Data Figure 3 of ref.~\cite{nature},
that however concern the total flux of 
neutrinos (i.e., all three flavors). 
The conclusion stated in Ref.~\cite{nature}, based on the observed $\gamma$-ray fluxes 
and on the criterion stated in \cite{vissanga}, 
is that these fluxes are potentially observable.
\newline
\indent
Here, we would like to proceed in the discussion further,  
clarifying the condition for observability 
in the existing detectors and
quantifying 
the expected number of signal events that can be detected. We will discuss how the conclusion depends upon the features of the detector.

\section{{Expected} signal in neutrino telescopes}
\label{sec:rates}

Current neutrino telescopes, like AN\-TA\-RES \cite{antares}, IceCube \cite{icecube} and those under construction as KM3NeT \cite{arca} and Bai\-kal\--GVD \cite{gvd}, could be able to detect the neutrinos from
the Galactic Center Region by looking for track-like events from the direction of this source. 

\paragraph{Track-{like} signal events and background events} 
The use of track-like events for the search of point sou\-rces 
is desirable because of the relatively good angular resolution, of the order of $1^\circ$ in ice and several times better in water. This allows the detectors to operate with a manageable rate of background events, due to atmospheric muons and neutrinos.

 The atmospheric neutrinos are an irreducible source of background events for all detectors. They can be discriminated from the signal of a point source due to the fact that they do not have a preferential direction, and moreover they have a softer energy spectrum
 than the one expected from the Galactic Center Region.
 {Part of the atmospheric neutrinos from above can be identified and excluded thanks to the accompanying muons, for  neutrino energies above 10 TeV and zenith angles  less than $60^\circ$ according to \cite{schonert}. 
This  rejection method works for the search of 
 High-Energy Starting Events (HESE) above 30 TeV in IceCube,  
 since it removes 70\% of atmospheric neutrinos 
in the Southern Hemisphere  \cite{icescience}.
Its application in our case is less effective. 
The first reason is obvious: 
Sgr~A* is observed at a high zenith angle from the South Pole, 
$\theta_Z=90^\circ-29^\circ=61^\circ$. 
Moreover, 
an important fraction of the signal is below 10 TeV,
as discussed later in this section.}

{
For what concerns atmospheric muons, one should 
distinguish, broadly speaking, the cases when the 
track-{type} events of interest for the search of the signal 
are upward-going or downward-going: \\
1)~The first class of events is not subject to the contamination of atmospheric muons. 
Due to the position of Sgr~A*, this kind of events can be observed by detectors located in the Northern hemisphere.\\
2)~The second class of events is subject to the contamination of atmospheric muons. 
Due to the position of Sgr~A*, this is relevant for 
IceCube. IceCube has successfully exploited a subset of downward-going track events with the  
purpose of investigating neutrino emission 
from Sgr~A* \cite{effI}, by requiring the additional condition 
that the production vertex of the downward-going tracks is contained in the detector.}

 



{To be precise, the fraction of time when the source is below the horizon is given by the expression 
$f_{\mbox{\tiny below}}=1- {\mbox{Re}[\arccos(-\tan\delta\tan\varphi)]}/{\pi}$ \cite{exa}. Its value is
\begin{equation}
f_{\mbox{\tiny below}}=0\%, 64\%,68\%,76\%
\end{equation}
for IceCube, KM3NeT-ARCA, ANTARES and GVD  respectively, where 
the declination of the Galactic Center is {equal to $\delta$(Sgr~A*)=-29.01$^{\circ}$}
and where the latitudes North of the various detectors are
$
\varphi= -90^\circ,36.27^\circ , 42.79^\circ , 51.83^\circ $  for IceCube, KM3NeT-AR\-CA, ANTARES and GVD respectively. 
In this fraction of time,  the  
atmospheric muon background is suppressed
and the search for a signal is easier.}

{The search for a signal with upward-going tracks allows  one 
to increase the effective volume of neutrino detection to the surrounding region, where the produced muon reaches the detector with sufficient energy.
Also the condition that the vertex is contained reduces the atmospheric muon background greatly, even in the low-energy region where it is more abundant.  
However, this condition does not allow to use the full volume of the detector but only a part of it, which hinders the search for a signal, especially a weak signal as the one we are discussing.

Another  specific circumstance favors the Che\-renkov telescopes operated in water, in comparison to those operated in ice, in the search for a neutrino signal at low energies. This is due to the angular resolution $\delta \theta$, that is better (i.e., smaller) in water than in ice. 
The number of background events $b$, falling in a given search window, decreases as  $\delta \theta^2$; the observable signal $s$ depends upon $s/\sqrt{b}$, that scales as $1/\delta \theta$.
{In any detector, there is a minimum energy below which the search for a signal becomes very challenging, since the number of background events tends to be excessively large. In water based detectors, this energy is  smaller than in the case of ice based detectors, simply because the number of background events decreases with angular resolution.}
For this reason, the  neutrino telescopes operated in water are more sensitive than the telescopes operated in ice, and can afford to use smaller energy threshold for data taking.

The IceCube {upper limit} mentioned in Sect.~\ref{sec:ripu} is based on  downgoing tracks and of course this telescope is operated in ice. We will show in the next paragraph that the use of a water based telescopes in the Northern hemisphere, able of good performances at low energies, can achieve significantly better results and have even the potential to probe the predictions of our model.}

{In principle, IceCube can also look at the Southern sky by exploiting the HESE sample \cite{icecube}, namely events of high energy with a vertex contained in the detector. The HESE are mainly composed of shower-type events, that have an angular {uncertainty} of about $10^\circ$, much worse than {that} of track-like events. Recall that 
IceCube used this data set to discover a population of diffuse (= unidentified) neutrino sources. 
The HESE sample was obtained adopting a very high energy threshold ($\sim 30$ TeV): this warrants a sufficiently clean sample, but requires a rather intense flux to produce an observable signal.
However, in the case of Sgr~A* we are {interested in} a point-source and to lower energy, so {a high angular} precision on the {reconstructed event} direction and an energy threshold much lower than 30 TeV {are needed}. We will show the importance of these considerations by a direct quantitative evaluation of the HESE event rate.} 

\begin{figure}[t]
\includegraphics[scale=0.49]{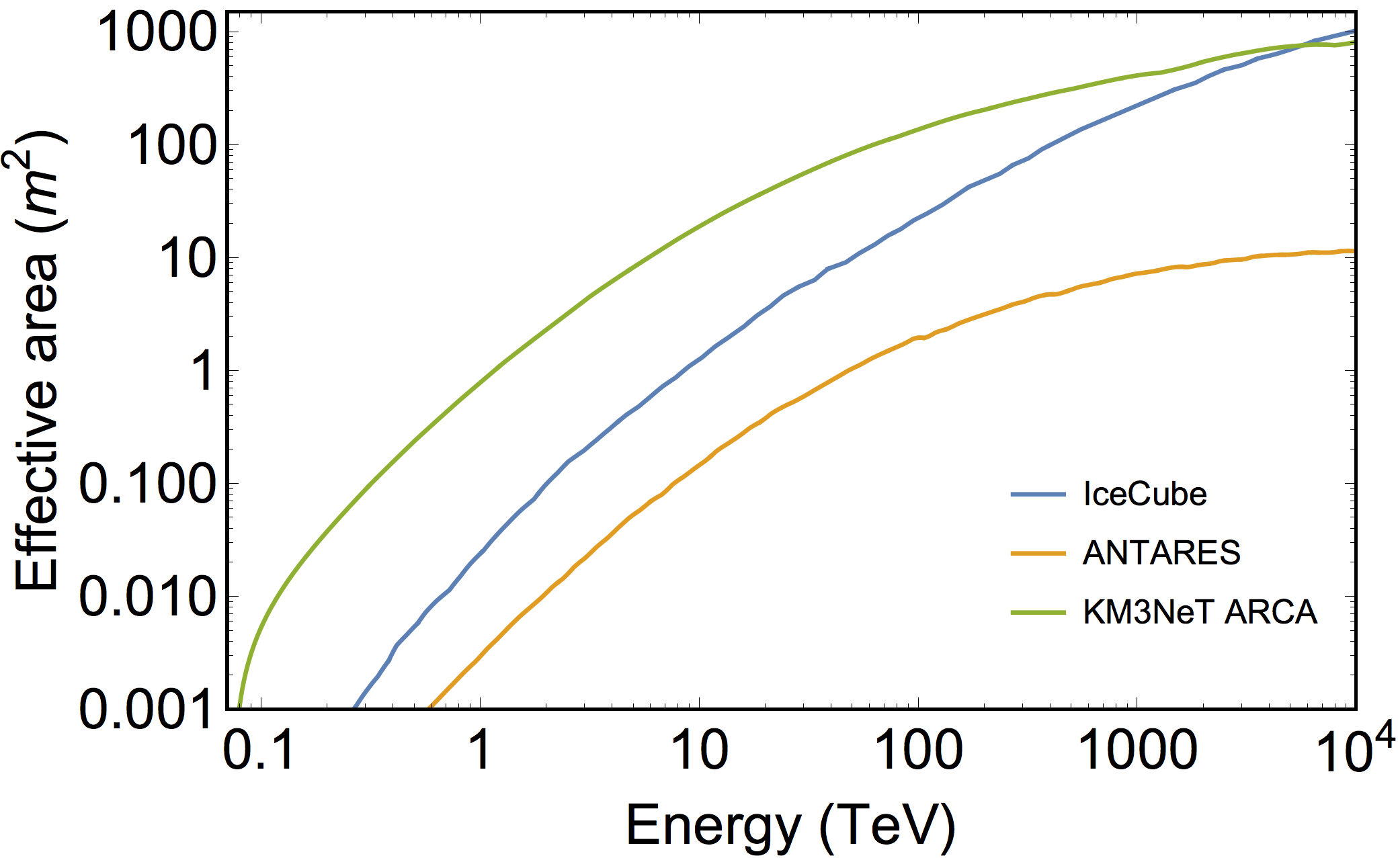}
\caption{{Muon neutrino effective areas for point-source search
adopted in the calculation. {The one adopted for ANTARES is that in the declination band $-45^{\circ}<\delta<0^{\circ}$, the one for IceCube is for $-30^{\circ}<\delta<0^{\circ}$, the one for ARCA is an average value}. 
See the text for references and for discussion.}}
\label{fig:aeff}
\end{figure}

\paragraph{Effective areas} 
The angular resolution for current neutrino telescopes is such that both the PS and the D regions are seen as point-like regions. 
Thus, the effective areas of ANTARES \cite{effA} and IceCube~\cite{effI} are those used for the search of point-like sources in the declination range relevant for the observation of the Galactic Center. 
Likewise, the effective area of KM3NeT-ARCA \cite{effK} {refers to the point source search: it is applied to the next configuration including two building blocks, each with 115 detection units.
These effective areas, that we use for the calculation of the rates, are shown in Fig.~\ref{fig:aeff}}.

\begin{figure*}[t]
\centering
\caption{{Parental neutrino energies of the signal in arbitrary units. From left to right the Point Source case (PS), the Diffuse case (D) and the Diffuse with a cut-off at {E$^{\gamma}_{\textrm{\tiny cut}}=600$}~TeV (DC). See also 
Tab.~\ref{tabparental}.}}
\includegraphics[scale=0.3]{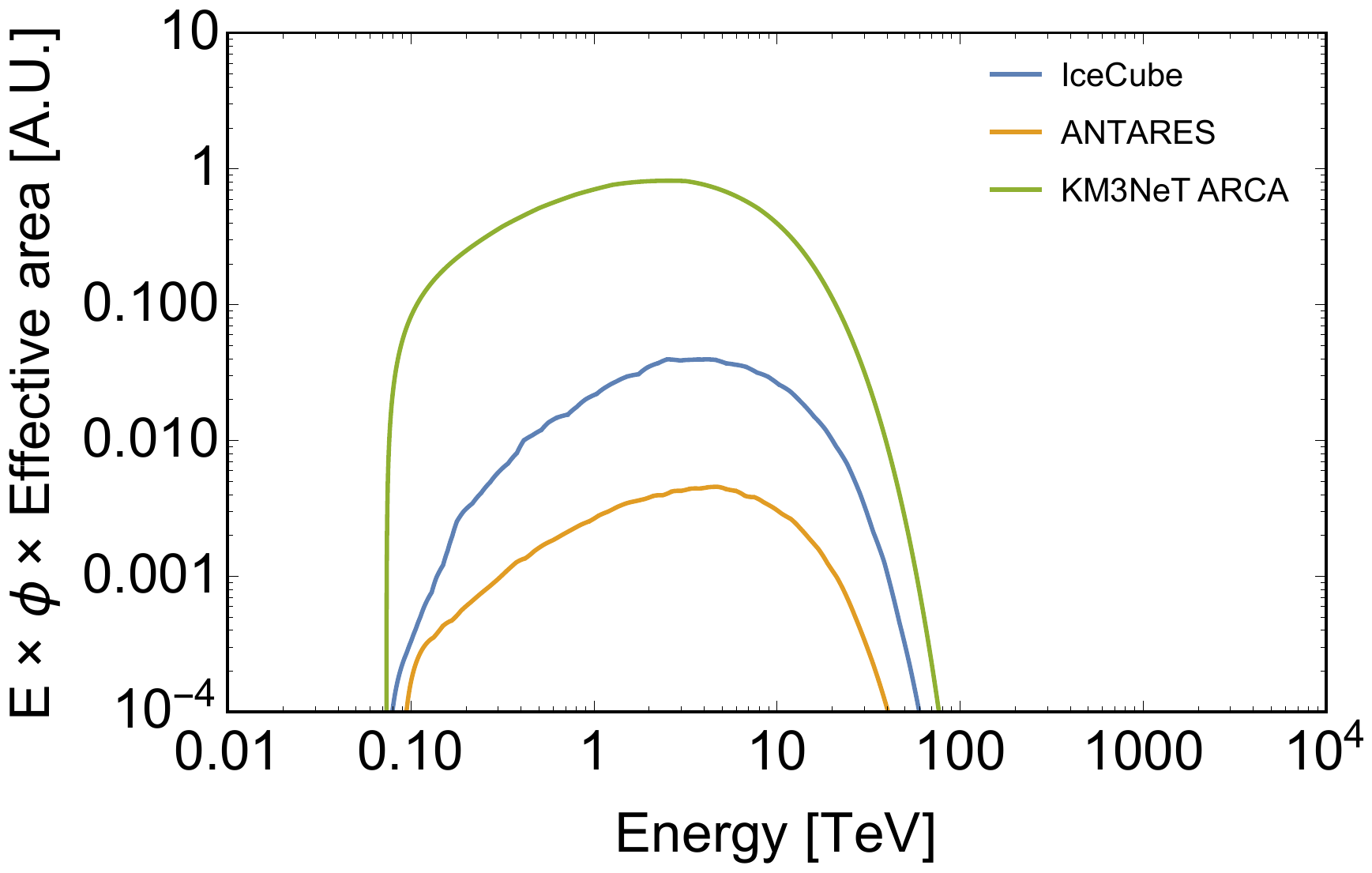}
\includegraphics[scale=0.3]{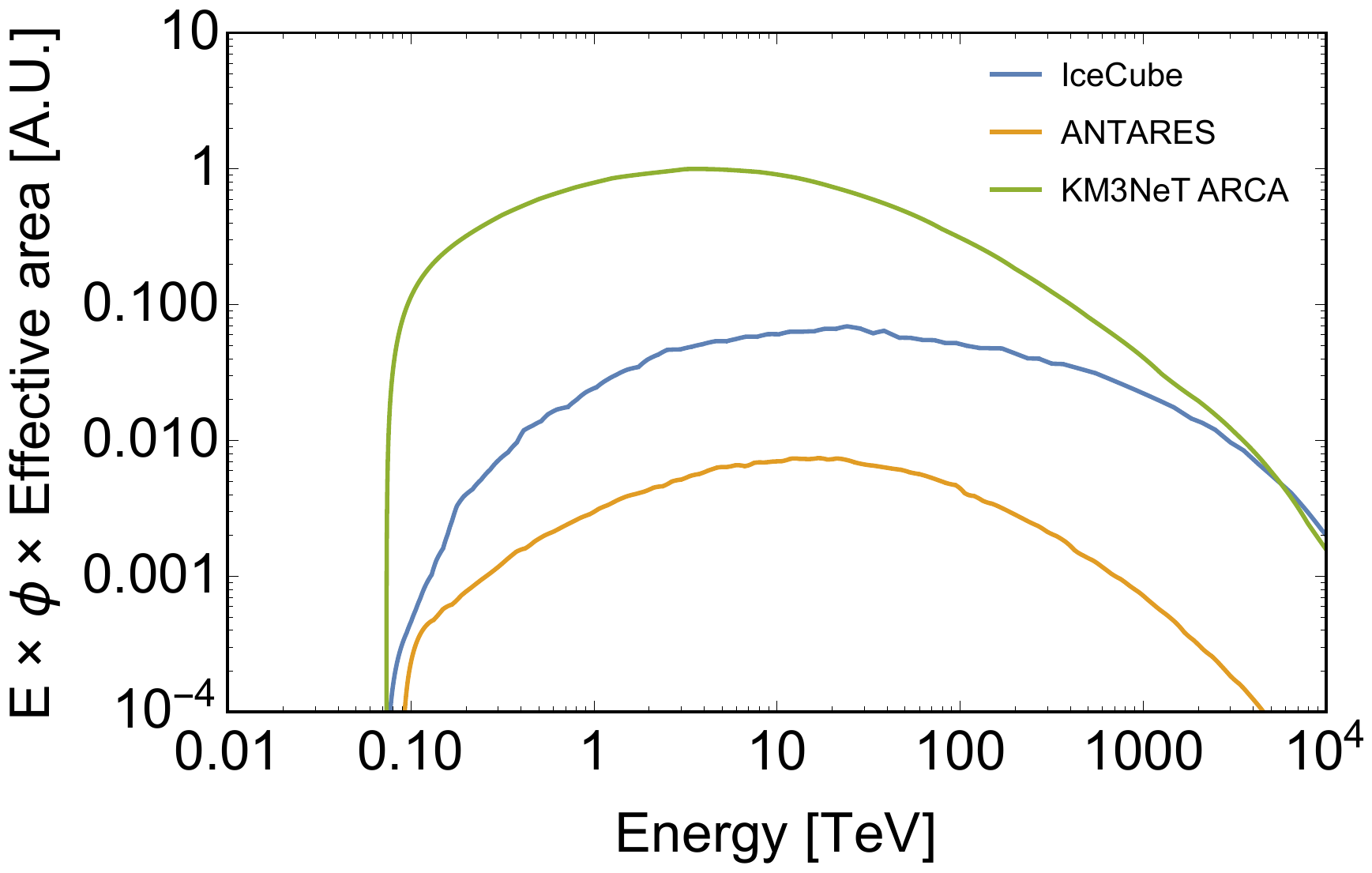}
\includegraphics[scale=0.3]{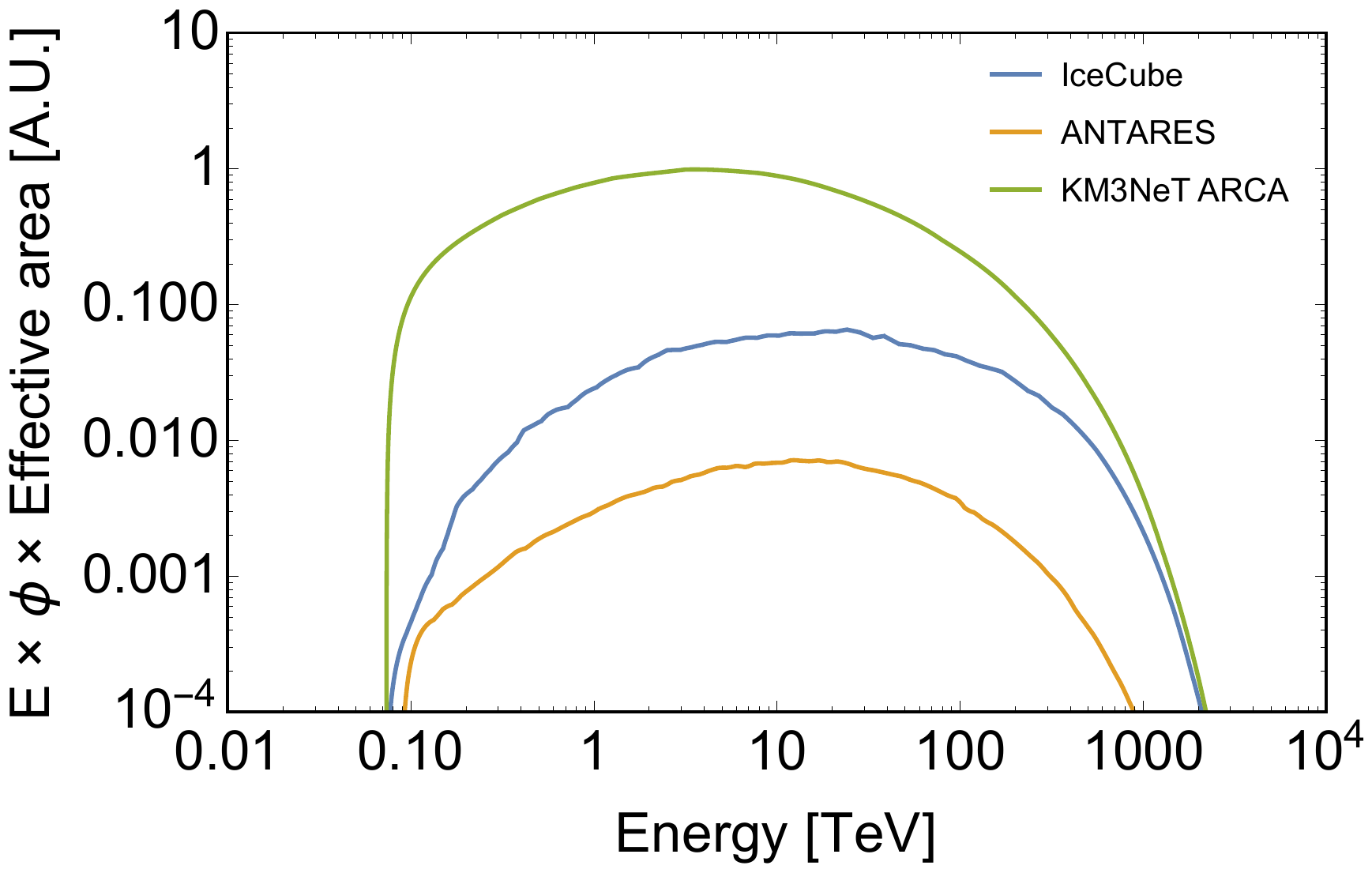}
\label{figpar}
\end{figure*}

The rate of events that a detector is able to measure, assuming a certain angular search region,  
is given by the convolution of the expected flux from the source, $\phi_{\nu_\mu}(E)+\phi_{\bar{\nu}_\mu}(E)$, and the detector effective area, $A_{\textrm{\tiny eff}}(E)$, through the relation,
\begin{equation}
\textrm{R}=\int{[\phi_{\nu_\mu}(E)+\phi_{\bar{\nu}_\mu}(E)]\, A_{\textrm{\tiny eff}}(E) \,dE}
\end{equation}
{The integrand in this formula, namely the product of the neutrino fluxes and of  the effective areas, 
is  the distribution of parent neutrino energies. 
Therefore, using the neutrino fluxes of Fig.~\ref{fig:fluxes},  we can evaluate 
the neutrino energies that contribute to the point source (PS) signal
for IceCube, ANTARES and KM3NeT-ARCA.
 The results are reported in Fig.~\ref{figpar} and in Tab.~\ref{tabparental}.
This proves that the  signal expected from Sgr~A* is at relatively  
low energies.}

\begin{table*}[t]
\caption{{Here reported are the median energy E$_{\tiny{50\%}}$ of the parental distribution and energy interval $[$E$_{\tiny{16\%}}-$E$_{\tiny{84\%}}]$ where 68\% of the signal is expected to be detected for the same models considered in 
Fig.~\ref{figpar}: the Point Source flux (PS), the Diffuse flux (D) and the Diffuse flux with Cut-off at E$^{\gamma}_{\tiny{cut}}=600$~TeV (DC). 
All energies are given in TeV.}}
\begin{center}
\begin{tabular}{c|ccc|ccc|ccc}
\hline
& & ANTARES & & & ARCA & & & IceCube &   \\
\hline
& E$_{\tiny{50\%}}$ & E$_{\tiny{16\%}}$ & E$_{\tiny{84\%}}$ & E$_{\tiny{50\%}}$ & E$_{\tiny{16\%}}$ & E$_{\tiny{84\%}}$ & E$_{\tiny{50\%}}$ & E$_{\tiny{16\%}}$ & E$_{\tiny{84\%}}$  \\
\hline
PS & 3.3 & 1.0 & 9.2 & 3.4 & 1.1 & 9.2 & 1.9  & 0.5 & 6.3 \\
D & 21.9  &  2.8 & 179.1 &  15.0  & 2.6 & 89.2 & 4.8  & 0.8 & 34.4 \\
DC & 14.8 &  2.3 & 88.5 &  12.1  & 2.3 & 60.1  & 4.3  & 0.8 & 26.3 \\
\hline
\end{tabular}
\end{center}
\label{tabparental}
\end{table*}%

{As discussed above, KM3NeT and ANTARES can look at the Galactic {C}enter using upward-going muons, while IceCube {has} to use downward-going muons further subject to the condition that their vertex is contained in the detector; moreover, the first type of 
telescopes {has} a much better angular resolution, which {allows them to reduce the background considered in the analysis, thus increasing the signal to noise ratio}. 
Despite the fact that the  dimensions of KM3NeT and IceCube are comparable, the effective areas
differ significantly {\em at low energies,} 
as can be ascertained from Fig.~\ref{fig:aeff}.
{Note however that the two effective areas become very similar around PeV energies, as expected because of the similar physical  sizes of these two neutrino telescopes.}
The different effective areas lead to the difference in the  number of events expected in IceCube and KM3NeT, which amounts to
about an order of magnitude.}

\paragraph{{Remarks}}

One cautionary remark on KM3NeT-ARCA effective area is in order. To the best of our knowledge, Ref.~\cite{effK} is the only public source of an effective area  for point source search with the KM3NeT-ARCA neutrino telescope.  We will use it to evaluate the expected signal from the Galactic Center Region, since this is the best that we can do at present, but if one wants to maintain a cautious attitude, one should contemplate the possibility that 
the experimental cuts adopted and consequently the 
effective  area will change in future releases. 
As we noted, the existing effective area  is quite large, e.g., in comparison to the one of IceCube, but as we demonstrate below, it corresponds to a signal of a few events per year only. Thus, it will be important to know whether the  experimental cuts, that will be eventually implemented by the KM3NeT {collaboration} for the search of the signal, will be compatible with similarly large effective area or will imply its revision.

{Finally, we would like to complete the discussion about the reason why IceCube cannot usefully 
exploit the data set at lower energies. 
Since the atmospheric neutrino background has a steeper spectrum with respect to the cosmic neutrino spectrum, at low energies it grows more than the signal, 
and therefore a very stringent selection has to be implemented on data in order to reject such a background. This can be for example realized through a tag: 
the atmospheric neutrino tag based on the accompanying muons 
works if the muons reach the detector with sufficient energy. 
The accompanying muons sho\-uld have enough energy in the 
production point to be revealed, which in turn means that the neutrino should have high energy, too. In our case, however, a significant part of  the signal is below 
the lower value of 10~TeV indicated in \cite{schonert}, as it is shown in Tab.~\ref{tabparental}. 
{This implies that  the efficiency of the atmospheric neutrino tag  
is less than for the search of  HESE
 above 30 TeV \cite{icescience}.
 Recall that Sgr~A* is observed at a 
 zenith angle  larger than  
 the lower value of $60^\circ$ indicated in  \cite{schonert},
and therefore, the accompanying 
muons lose a significant amount 
of energy before reaching IceCube.}\footnote{IceCube is 
at a depth of 1.45 km$<h<$2.82 km; thus,  
muons pass a relatively large amount of ice, 
$h/\cos\theta_Z\approx 2\times h$.} 
These considerations limit to relatively high energies 
the region where IceCube may conveniently search for a 
point source from Sgr~A*, as quantified by the effective area reported by the IceCube collaboration. }


\begin{table*}[t]
\caption{{\em First 4 columns:}  
Spectral parameters {assumed for the} $\gamma$-ray {fluxes, consistent with the H.E.S.S. observations as explained in the text}: the search region (PS=Point Source or D=Diffuse), the spectral index $\Gamma$, the flux normalization $\phi_0$ in units of 10$^{-12}$ TeV$^{-1}$ cm$^{-2}$ s$^{-1}$ and the energy cut-off E$_{\textrm{\tiny cut}}$ in TeV (see Eq.~\ref{eq:plcut}). 
For the PS and the D models, 
we show also the maximum and minimum expected values. 
{For the PS* model we assume a scenario with an increased, 
non standard $\gamma$-ray absorption.}
{\em Last 4 columns:} 
Expected number of $\nu_\mu+\bar\nu_\mu$ events  per year: 
{downward-going tracks and HESE events in IceCube, 
upward-going tracks in ANTARES and ARCA.}
The significant increase of the event rate passing from the 
PS (1$^{st}$ row) to the PS* (4$^{th}$ one) model is linked to the non standard $\gamma$-ray absorption.}
\begin{center}
{
\begin{tabular}{c|ccc|cccc}
\hline
& & $\!\!\gamma$-rays &  & & $\!\!\!\nu_\mu+\bar{\nu}_\mu\!\!\!$ \\
\hline
 & $\Gamma$ & $\phi_0\!$ & E$_{\textrm{\tiny cut}}\!$ &
 R$^{\textrm{\tiny ANTARES}}\!\!$ & 
 R$^{\textrm{\tiny ARCA}}\!\!\!$ & 
 R$^{\textrm{\tiny IC}}\!\!$ &
R$^{\textrm{\tiny IC}}\!\!_{\textrm{\tiny HESE}}$ \\
\hline 
\small{PS} & \small{2.14} & 2.55 & 10.7 & $6.2 \cdot 10^{-3}$ & 1.1 & $5.2 \cdot 10^{-2}$ & $1.4 \cdot 10^{-6}$\\
''& 2.04 & 2.92 & 13.6 & $9.5 \cdot 10^{-3}$ & 1.5 & $8.2 \cdot 10^{-2}$ & $6.1 \cdot 10^{-6}$ \\
'' & 2.24 & 2.18 & 7.8 & $3.9 \cdot 10^{-3}$ & 0.7 & $3.2 \cdot 10^{-2}$ & $1.9 \cdot 10^{-7}$ \\
\small{PS*} & 2.14 & 2.55 & 100 & $1.7 \cdot 10^{-2}$ & {2.1} & $1.5 \cdot 10^{-1}$ & $5.0 \cdot 10^{-4}$\\ \hline
\small{D} & 2.32 & 1.92 & - & $1.2 \cdot 10^{-2}$ & 1.4 & $1.3 \cdot 10^{-1}$ & $2.2 \cdot 10^{-3}$\\
'' & 2.20 & 2.21 & - & $2.1 \cdot 10^{-2}$ & 2.2 & $2.6 \cdot 10^{-1}$ & $5.5 \cdot 10^{-3}$\\
'' & 2.44 & 1.63 & - & $7.5 \cdot 10^{-3}$ & 1.0 & $7.4 \cdot 10^{-2}$ & $8.8 \cdot 10^{-4}$\\
\small{DC} & 2.32 & 1.92 & 400 & $1.0 \cdot 10^{-2}$ & 1.3 & $9.7 \cdot 10^{-2}$ & $6.8 \cdot 10^{-4}$ \\
\small{DC} & 2.32 & 1.92 & 600 & $1.1 \cdot 10^{-2}$ & 1.3 & $1.0 \cdot 10^{-1}$ & $8.8 \cdot 10^{-4}$ \\
\small{DC} & 2.32 & 1.92 & 2900 & $1.2 \cdot 10^{-2}$ & 1.4 & $1.2 \cdot 10^{-1}$ & $1.6 \cdot 10^{-3}$\\
\hline
\end{tabular}}
\end{center}
\label{tab:nuExp} 
\end{table*}%

\paragraph{Expected signal rates} The rates of events for ANTARES, KM3NeT-ARCA and IceCube are given in Tab.~\ref{tab:nuExp} with the names $\mbox{R}^{\tiny \mbox{ANTARES}}$, $\mbox{R}^{\tiny \mbox{ARCA}}$, $\mbox{R}^{\tiny \mbox{IC}}$,
considering the different spectral model of $\gamma$-ray data presented above, and accounting for both contributions from muon neutrinos and antineutrinos, as expressed in Eq.~\ref{eq:Inu}.
Baikal-GVD will have a
threshold of few TeV and a volume similar to 
KM3NeT-ARCA, so the results are expected to be similar, 
but we cannot provide a precise evaluation of the signal since we do not have the effective area.

For comparison, we calculated that the expected rate corresponding to Eq.~\ref{ict} (namely assuming a E$^{-2}$ distribution) in IceCube 
is 3.8 per year, namely, more than one order of magnitude above the values of  Tab.~\ref{tab:nuExp}. This illustrates the great importance of investigating the $\gamma$-ray distribution at higher energies than currently observed.

As can be seen from Tab.~\ref{tab:nuExp}, the detectors located in the Northern hemisphere are better suited for neutrino searches from Sgr~A*.
In fact, when the source is below the horizon, 
they can observe the Galactic Center Region through upward-going track events, that are not polluted by the atmospheric 
muons.  
Such detectors are ANTARES,  Baikal-GVD and KM3NeT. We find that the expected rates in AN\-TARES are just one order of magnitude smaller than those expected in IceCube with downward-going events: this result is well in agreement with the ones in \cite{spurio}.  

{For completeness, the rates of the expected HESE 
track events are also given in Tab.~\ref{tab:nuExp}. The counting 
rate is indicated with the name of $\mbox{R}^{\tiny \mbox{IC}}_{\tiny \mbox{\ \ \ HESE}}$ and it was 
obtained 
using the effective areas reported in \cite{Aartsen:2013jdh}.
The counting rates are, in the best case, 
few times $10^{-3}$ HESE events per year.
Therefore,  this approach does not allow IceCube   
to search for neutrinos from the Galactic {C}enter.}


\paragraph{Discussion} Among the $\gamma$-ray models presented in this table, the most plausible ones are, presumably, those described by a power law {\em with} a cut-off. 

In the Diffuse case, even considering the less favorable case (the one with lowest energy cut-off, which implies a cut off in the primary spectrum of protons at about 4 PeV, {where the \textit{knee} of the Earth-observed CR spectrum is located}) predictions are such that the incoming km$^3$ class detectors in the Northern hemisphere as KM3NeT-ARCA could measure these neutrinos with a rate of few events per year: several years of data-taking will be in any case needed in order to establish the presence of a proton galactic accelerator up to PeV energies and address the origin of very-high-energy cosmic rays. {In case of non-detection, however, strong constraints will be derived concerning the proton acceleration efficiency of this poorly-understood source.}

A similar conclusion holds  true for the Point Source {case}. When we assume that the cut-off measured by H.E.S.S.\ is due to the absorption by a non standard background radiation field, the muon neutrino signal increases. E.g., comparing the first row (PS case) and the fourth one (PS*) of Tab.~\ref{tab:nuExp} we see an increase by 40-50\%; note that the parameter of the exponential cut-off  has been 
set to 100 TeV in the PS* case.
Remarkably, the protons accelerated in the source can reach energies up to the PeV scale in this scenario.

Unfortunately, with the current neutrino telescopes,
these predictions cannot be probed yet. {Anyway, since most of the signal is expected in the 1-100 TeV energy band, the Northern hemisphere telescopes cannot escape from the issue of atmospheric neutrino background events.}

\section{Conclusions\label{sec:ccc}} 
The recent measurements  of multi-TeV $\gamma$-rays from the Gala\-ctic Center, performed by H.E.S.S., point out the chance of observing  also very-high-energy neutrinos from this part of the Galaxy. The detection of such neutrinos is crucial to confirm or discard the hadronic origin of these $\gamma$-rays. {In case of non-detection, however, neutrino telescope will be able to put severe constraints on the efficiency of hadronic acceleration in this source.}

In order to estimate the neutrino flux, it is necessary to know the flux of $\gamma$-rays at the source, and for this reason it is crucial to evaluate correctly the effect of the absorption due to the interaction between $\gamma$-rays and the background radiation fields. 
 We have argued that the Diffuse high-energy $\gamma$-ray flux measured by H.E.S.S.\ is not affected by the absorption. 
 On the contrary, the $\gamma$-ray flux from the Point Source could be affected by an intense infrared radiation field that, if it exists, should be present close to the Galactic Center. {We have found that this effect is compatible with an unbroken power law distribution and  
can increase the observable neutrino signal by 40-50\%.} 
 This new hypothesis motivates further studies with IR telescopes and with 100 TeV $\gamma$-ray instruments, as the future Cherenkov Telescope Array~\cite{cta}.

We have obtained  a precise upper limit on the expected neutrino flux from the regions close to the Galactic Center, 
assuming that the $\gamma$-rays  
recently observed by H.E.S.S.\  
are produced by cosmic ray collisions.
As shown in Tab.~\ref{tab:nuExp}, the  corresponding maximum signal is of few {track (muon signal) events} per year 
in the incoming KM3\-NeT detector. {In view of these results, we conclude that the KM3NeT detector has the best chances to observe neutrinos from Sgr A*,
even if, in order to accumulate a large sample of signal events, several years of exposure will be necessary. 
Besides the analyses of the track-like events in KM3NeT, discussed above, also the analyses of shower-like events will contribute to advance the study of Sgr A*, thanks  to the favorable location of this detector and to its superior angular resolution.}
On the contrary the expected signal in  IceCube is smaller and unlikely to be observed in view of the larger background rate 
{caused by the atmospheric muons.}

We have examined the reasons of uncertainties in the expectations for the high-energy neutrinos. While a leptonic component of the $\gamma$-ray would  decrease the observable signal, several other reasons could increase it, including: the possibility of $\gamma$-ray absorption, an extended angular region around the Galactic Center where the emission is sizeable, 
a speculative E$^{-2}$ behavior of the spectra at higher energies than presently measured with $\gamma$-rays.  
These considerations emphasize the importance of extending the programs of search and study of high-energy $\gamma$-rays and neutrinos.

\section{Acknowledgments}
We thank F.~Aharonian,
S.~Gabici, F.~Fiore, A.~Lamastra  and F.~Tavecchio 
for precious discussions. 
After this study was concluded, two other investigations of 
$\gamma$-ray absorption
appeared:
Ref. \cite{guo} draws a similar picture 
concerning non standard absorption, Ref.~\cite{pl} focusses on 
standard effects instead, and 
our Fig.~\ref{figabs} and their figure  12 are in perfect 
agreement.

\appendix
\section{Study of the function $f(x)$}

The evaluation of the effects of absorption via Eq.~(\ref{tau}) rests on the estimation of the properties of the background radiation field and on the calculation of a single universal function. 
Note in passing that, even if we are interested to use these results to $\gamma$-rays emitted from the Galactic Center, the results concerning $\gamma$-ray absorption can be applied to a very large variety of cases and situations.

In this appendix, we study $f(x)$ in details, providing a table of (virtually exact) numerical values for this function and discussing the bases of the approximation given in Eq.~(\ref{eq:analyticApp}). The following material is useful to verify our results and to compare with other results in the literature, but has been confined in this appendix, so that it can be skipped by the uninterested Reader.

The 
pair creation process \cite{breit} 
$$\gamma+\gamma_{\textrm{\tiny bkg}} \longrightarrow e^+ + e^-$$ 
in the background of thermal photons with temperature 
$T_i$
gives the opacity factor \cite{moska},
$$
\tau_i=\frac{1}{\pi}\times r_e^2 L_i\times T_i^3\times f\left(\displaystyle
\frac{m_e^2}{T_i\ E_\gamma}\right)
$$
that can be rewritten introducing the thermal photon density 
$n_{\gamma,i}=2 \zeta(3) T_i^3/\pi^2$.
The function $f(x)$ is defined as,
\begin{equation}
f(x)=x^2 \int_0^1 \! d\beta\ R(\beta)\ \psi \left(\frac{x}{1-\beta^2}\right) 
\label{unifun} 
\end{equation} 
Here $\beta$ is the velocity of the outgoing electron in the center of mass frame, and the two auxiliary functions are,
$$
\begin{array}{l}
R(\beta)=\displaystyle\frac{2\beta}{(1-\beta^2)^2} \left[ (3-\beta^4) \log\left(\frac{1+\beta}{1-\beta}\right) - 2\beta(2-\beta^2)\right]  \\[2ex]
\psi(z) =\displaystyle-\log\left(1-e^{ -z} \right)
\end{array}
$$ 
with $z=x/(1-\beta^2)$.
Solving numerically the integral in Eq.~\ref{unifun} we found the values reported in Tab.~\ref{tab:val}.

\begin{table}[t]
\caption{Table of values of the function $f(x)$ as given in Eq.~\ref{unifun}. In bold, the value of $x$ in which the function reaches the maximum and half of the maximum.}
\begin{center}
\footnotesize
\begin{tabular}{cc|cc}
\hline
$x$ & $f(x)$ & $x$ &$f(x)$ \\
\hline
$10^{-10}$ & $7.32 \times 10^{-9}$ & $10^{-1}$ & $6.32 \times 10^{-1} $ \\
$10^{-9}$ & $6.57 \times 10^{-8} $ & \textbf{0.503} & \textbf{1.076} \\
$10^{-8}$ & $5.81 \times 10^{-7} $ & 1 & $9.07 \times 10^{-1} $ \\
$10^{-7}$ & $5.05 \times 10^{-6} $ & \textbf{1.77} & \textbf{0.538} \\
$10^{-6}$ & $4.29 \times 10^{-5} $ & 5 & $3.27 \times 10^{-2} $   \\
$10^{-5}$ & $3.54 \times 10^{-4} $ & 10 & $2.92 \times 10^{-4} $  \\
$10^{-4}$ & $2.78 \times 10^{-3} $ & 20 & $1.78 \times 10^{-8} $ \\
$10^{-3}$ & $2.04 \times 10^{-2} $ & 30 & $9.65 \times 10^{-13} $ \\
$10^{-2}$ & $1.31 \times 10^{-1} $ & 50 & $\simeq  0$  \\
\textbf{0.0756} & \textbf{0.538} & & \\
\hline
\end{tabular}
\end{center}
\label{tab:val}
\end{table}

First, we examine the behaviour of the integrand in $\beta$. The function $R(\beta) \sim 4 \beta^2$ if $\beta \rightarrow 0$; on the contrary, when $\beta \rightarrow 1$, it diverges like $R(\beta)\sim -\frac{\log(1-\beta)}{(1-\beta^2)^2}$. The divergence is compensated by the behavior of the function $\psi(\frac{x}{1-\beta^2})$, that follows from $\psi(\frac{x}{1-\beta^2})=\sum_{n=1}^\infty \frac{e^{-n\frac{x}{1-\beta^2}}}{n}$ at high values of $\frac{x}{1-\beta^2}$. Finally,  $\psi(\frac{x}{1-\beta^2})\approx - \log \frac{x}{1-\beta^2}$ at small values of $\frac{x}{1-\beta^2}$.

At this point, we study the behaviour  of $f(x)$ in $x$: \\
For high $x$ we can consider only the first term of the expansion of $\psi(\frac{x}{1-\beta^2})$, so the function $f(x)$ is well approximated by:
$
f(x)\approx x^2  \times 
\int_0^1\ d\beta \ R(\beta) \exp \left(-\frac{x}{1-\beta^2} \right) 
$
within an accuracy of 1\% for $x>3$. \\
For small $x$ the most important contribution to the integral is given when the $R(\beta)$ diverges and the $\psi(\frac{x}{1-\beta^2})$ is not exponentially suppressed. This condition is realized when $\beta< \sqrt{1-x}\approx 1-x/2$ and in this region $\psi(\frac{x}{1-\beta^2}) \approx -\log \frac{x}{1-\beta^2}$; for $R(\beta)$ we can use the asymptotic expression, i.e. $R(\beta) \simeq 4 \frac{\log\big(\frac{2}{1-\beta} \big)}{(1-\beta^2)^2}$. The approximation of the function $f(x$) is given by:
$
f(x) \approx
-4x^2 \int_0^{1-x/2} $ $d\beta \frac{\log\big(\frac{2}{1-\beta} \big) \log \big(\frac{x}{1-\beta^2} \big)}{(1-\beta^2)^2} $. 
This implies the behavior, $f(x) \approx -3.076 \ x \ \log(x)$ to within an accuracy of about 3\% in the interval $10^{-10} \leq x \leq 10^{-5}$.  

A global analytical approximation of the $f(x)$, that respects the behavior for small and large values of $x$, is  given by
Eq.~\ref{eq:analyticApp}. Its accuracy is $\sim 3\%$ into the interval $10^{-10} \leq x \leq 10$. When $x > 10$ the function rapidly decreases, as we can see also from Tab.~\ref{tab:val}, where the values are obtained by numerically integrations without any approximation. 



 \end{document}